\pdfoutput=1
\documentclass{JINST}
\let\ifpdf\relax
\title{RPC Gap Production and Performance for CMS RE4 Upgrade}

\author{Sung Keun Park\thanks{Corresponding author.} , Min Ho Kang and Kyong Sei Lee\\
\\
\\
Korea Detector Laboratory, Department of Physics, Korea University, Seoul, 136-713,
Korea\\
\\
\\
E-mail: \email{sungpark@korea.ac.kr}}

\abstract{
CMS experiment constructed the fourth Resistive Plate Chamber (RPC) trigger station composed of 144 RPCs to enhance the high momentum muon trigger efficiency at both endcap regions. All new CMS endcap RPC gaps are produced in accordance with QA and QC at the Korea Detector Laboratory (KODEL) in Korea. All qualified gaps have been delivered to three assembly sites: CERN in Switzerland, BARC in India, and Ghent University in Belgium for the RPC detector assembly. In this paper,  we present the detailed procedures used in the production of RPC gaps adopted for the CMS upgrade.
}


\begin{document}

\section{Introduction}



In the current project, a total of 144 RE4 Resistive Plate Chambers (RPCs) are manufactured, tested, and are being installed at the fourth endcap RPC station in the CMS detector.  One unit of the standard CMS endcap RPCs consists of 3 gaps forming double layers. It means that 432 gas gaps should be produced for the assembly of 144 RE4 RPC modules. An additional 200 gaps are produced for the contingency and future repairs of any damaged RPC modules. 

The RE4 RPCs are required to be mounted in two concentric rings RE4/2 and RE4/3 with 36 RPCs per ring. The shapes of the gaps are trapezoidal. The heights of the RE4/2 and RE4/3 gaps are 1663 and 1930 mm, respectively~\cite{cmsrpc}.

The Korea Detector Laboratory (KODEL) produces the endcap RPC gaps for CMS. KODEL utilizes the dedicated RPC mass production facilities for gaps in which the entire CMS endcap RPC gaps have been produced since 1997.

The RPC gaps have been delivered to three RPC assembly sites; CERN in Switzerland, BARC in India, and Ghent University in Belgium for the detector assembly. 






\section{Gap Production}


\begin{figure} [h]
\begin{center}
\includegraphics[width=9.0cm]{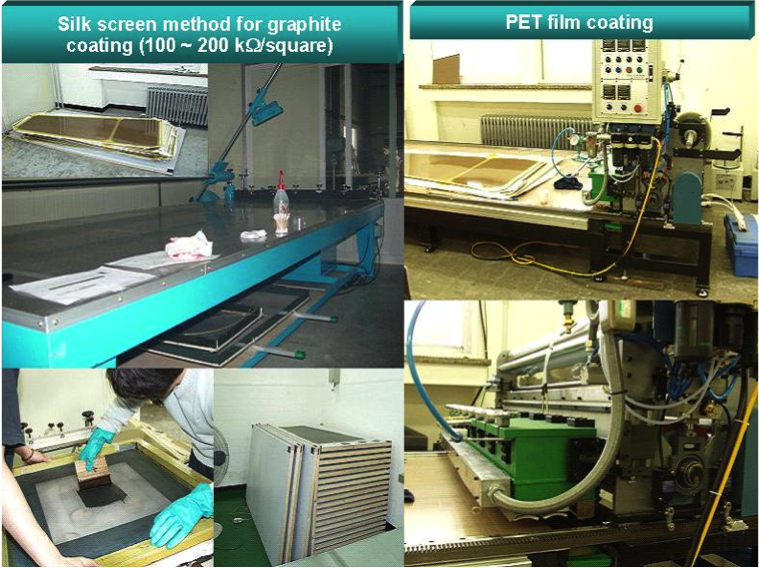}
\caption{Graphite coating facility (left column) and PET film coating facility (right column).}
\end{center}
\label{fig:coating}
\end{figure}



Mass production of the RPC gaps relies on several major facilities at KODEL: the graphite coating facility, the PET film coating facility,
the spacer bonding and gas leak test facility, the linseed oil treatment facility,
and high voltage (H.V.) test facility where dark currents are monitored and measured.

Quality Control (QC) is imbedded in the process of RPC gap production. Because QC can be defined as a collection of protocols to select the product that meets the technical specifications, all of the produced gaps are subject to QC.

QC is further divided into the components of the RPCs. The main item for QC-1 is High Pressure Laminate (HPL) panels, the Bakelite. QC-2 is for gaps while QC-3 is for chamber assembly and operation. QC-3-1 concerns chamber assembly and QC-3-2 concerns performance assessment of the detector with cosmic ray muons. QC-4 is for chambers and super modules.
QC in this paper will be focused on the QC-2 designed for the gap production.

\subsection{Gas Volume Production}
%


\begin{figure} [h]
\begin{center}
\includegraphics[width=9.0cm]{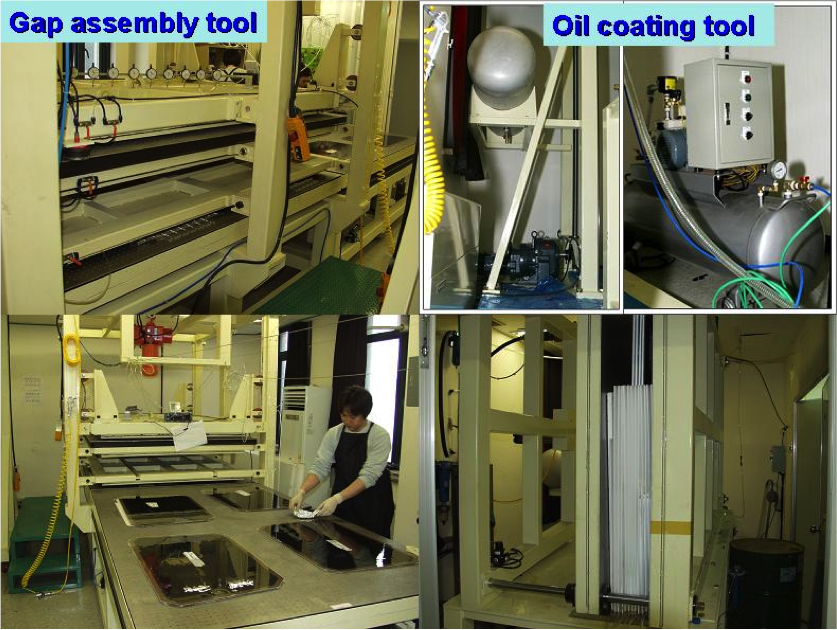}
\caption{ Gap bonding facility (left column) and linseed oil treatment facility (right column).}
\end{center}
\label{fig:bonding}
\end{figure}

The electrodes of the RPC gaps are produced at KODEL with a thin layer of graphite.
The surface resistivity of the electrode is checked as the first step of the QC-2.
The surface resistivity should range from 100 to 250~k$\Omega$/square~\cite{surface}.

\begin{figure}[ht]
\begin{center}$
\begin{array}{cc}
\includegraphics[width=7cm]{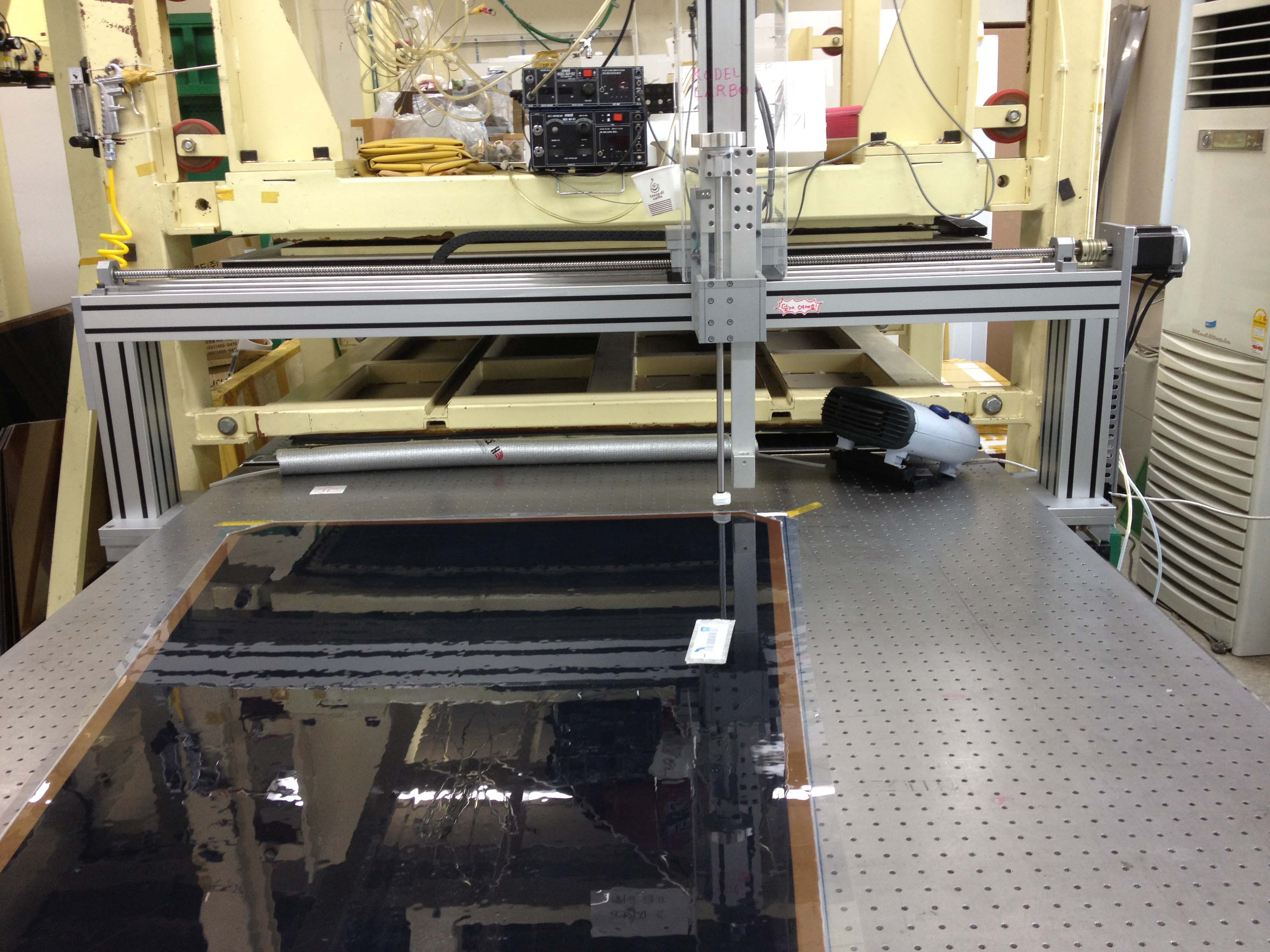} &
\includegraphics[width=7cm]{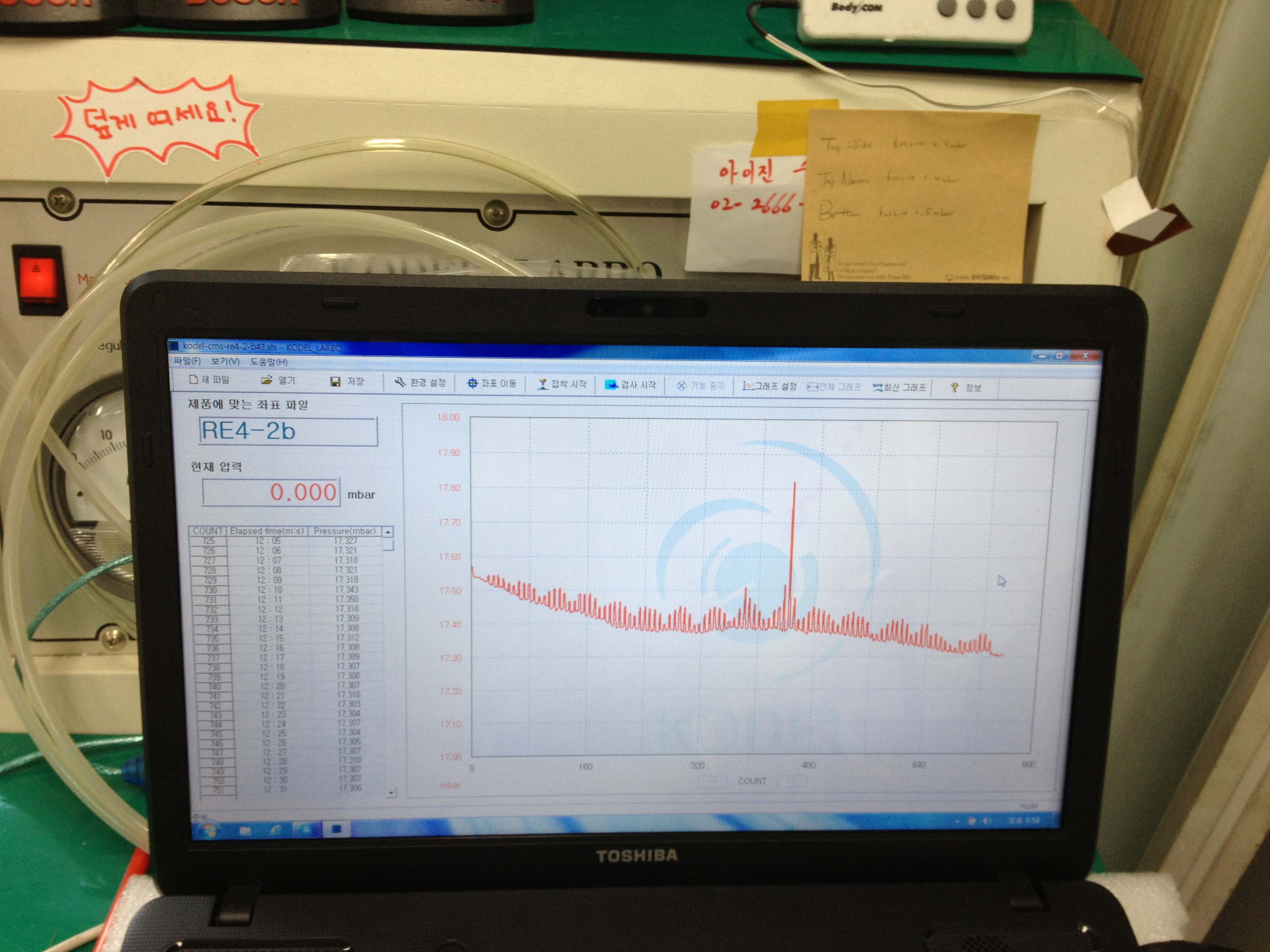}
\end{array}$
\end{center}
\caption{KODEL robot performing the pressure test. The dimensions of the table are 175 cm x 250 cm (left). The control console of the pressure test (right). }
\label{fig:robot} 
\end{figure}

The next step of the QC-2 is to check the mechanical stability of the gap once it is produced. To secure the uniform electric field within the gap, homogeneous 2 mm separation is achieved by the spacers. The secure bonding of the spacers on the surface of the electrode made of Bakelite is checked. The pressure variation of the gaps when the spacer is pressed by a KODEL robot is an effective indicator of the good bonding, as Fig.~\ref{fig:robot} shows.

Another property of the RPC gap to check in QC-2 is the gas leakage rate. This leakage rate measurement is performed within the frame work of the bonding test.

Fig. 4 shows the results of QC-2 registering the pressure variation when pressed on the spacers. 
Differences at the beginning and end of the pressure measurement include the gas leak rate of the gas gaps. The curve usually shows a gradual decreases in pressure if the gap has a small leakage rate.

If a gap without loosen spacers has the gradual slope of the gas leak within the acceptable range, the gap is prepared to move to a linseed-oil procedure. The allowed maximum leakage rates for 10 minutes are 0.2, 0.3 and 0.4 hPa for RE4-2 Top Wide (TW), Top Narrow (TN) and Bottom gaps, respectively. For RE4-3 gaps 0.3, 0.4, 0.5 hPa for TW, TN and Bottom gaps, respectively.

\begin{figure}[ht]
\begin{center}$
\begin{array}{cc}
\includegraphics[width=7cm]{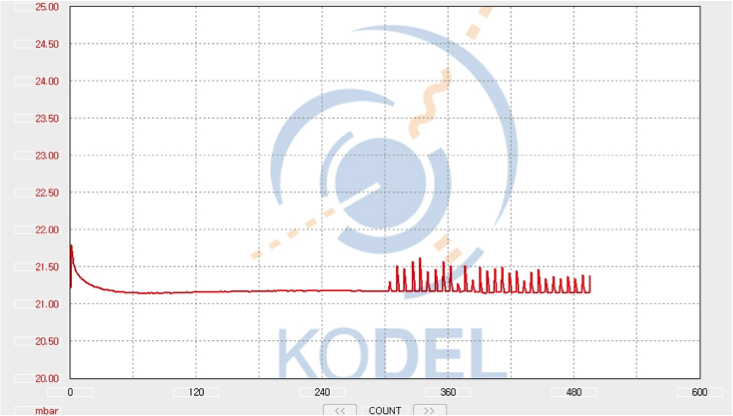} &
\includegraphics[width=7cm]{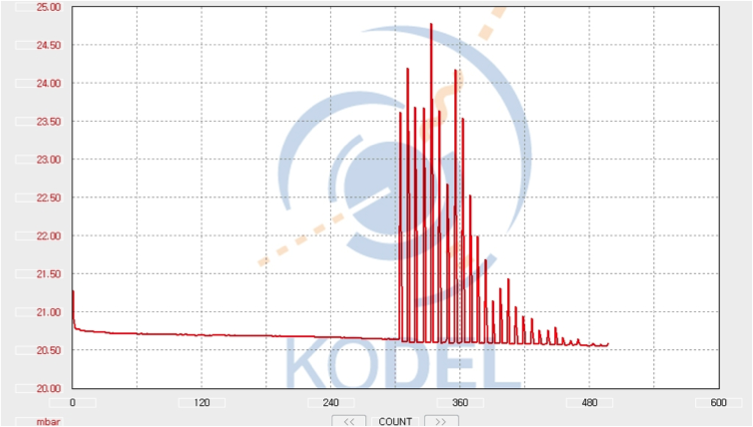}
\end{array}$
\end{center}
\caption{Outcome of the pressure test. Pressure variation of the gap for all secured spacers (left) and loosen spacers (right). }
\end{figure}

\subsection{Linseed Oil Treatment}

During the linseed oil treatment process, the entire gas volume is filled with linseed oil. Once dried with air, the oiled surface forms the polymerized layer. 
It shows that the polymerized layer on the surface of the bakelite inside the gas volume reduces the spurious noise by a factor of 10 in the avalanche operation mode~\cite{kodelrpc2}. This procedure needs a cautious preparation due to the pressure built up by the linseed oil filling the gas gap. The pressure within the gap should remain less than 20 hPa over the atmospheric pressure to ensure that the gas gap is not over-pressured to the limit of its epoxy bonding strength. 


\subsection{Dark Current Measurement}
 


\begin{figure} [h]
\begin{center}
\includegraphics[width=9.0cm]{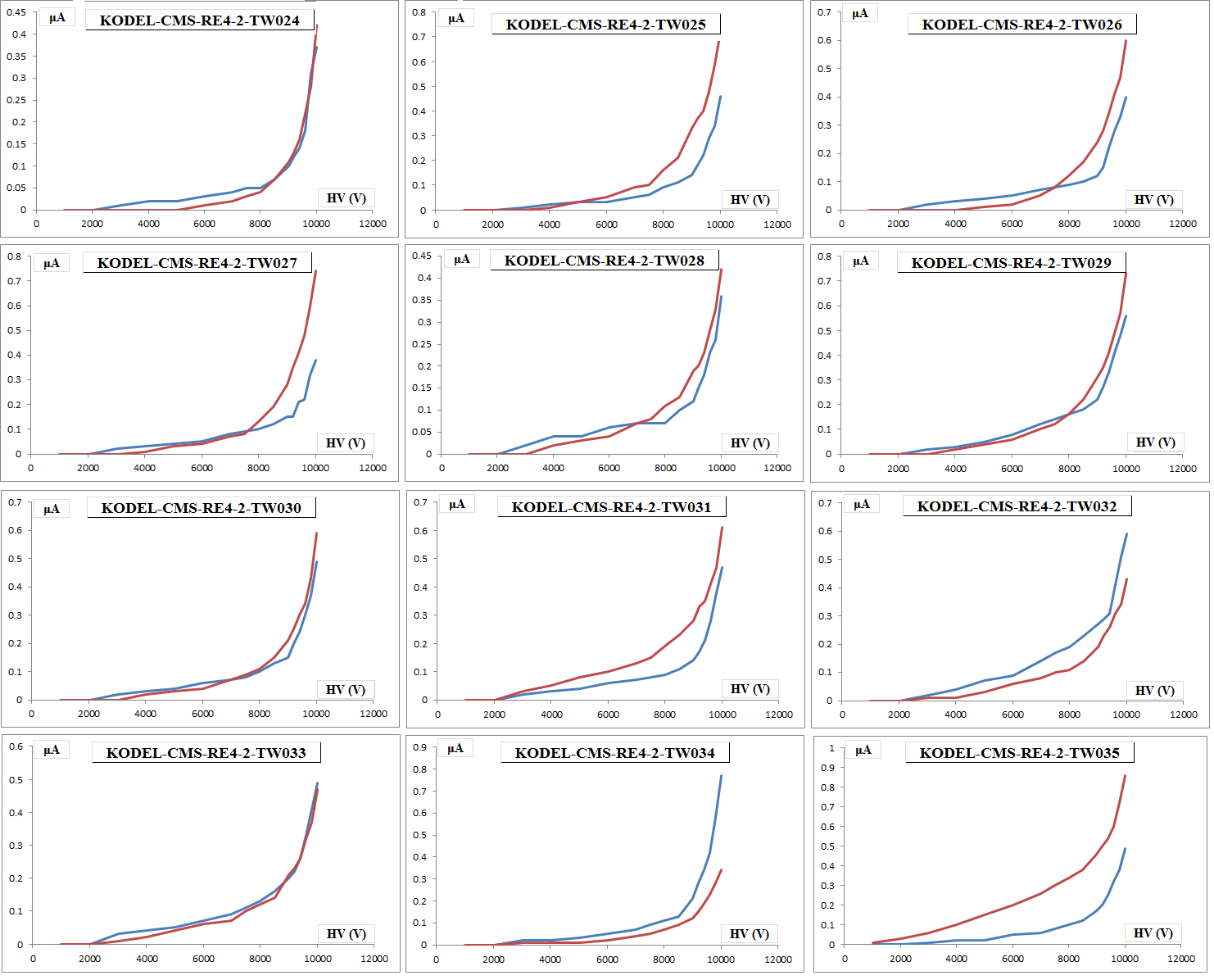}
\caption{Dark currents of gaps as a function of H.V. up to 10 kV before (blue) and after 120-hour long test (red). }
\label{fig:dark_current10}
\end{center}
\end{figure}

The dark current measurement has three steps. The first part is a H.V. scanning up to 10 kV. Dark current is measured as H.V. increases up to 10 kV as shown
 in Fig.~\ref{fig:dark_current10}. In this step, the amount of current drawn from gaps at 6 kV should not exceed 1.5~$\mu$A. At 10 kV, the maximum 
 5~$\mu$A is allowed for smaller gaps and the maximum 10~$\mu$A for larger gaps. 

Each plot has two components: one linear component rising from zero up to 7 kV and another with an exponential rise from 8 kV up to 10 kV. It is the characteristic curve in the HV scanning. Dark current at 6 kV is important; at this point the current is not originated from the avalanche of the gas ionization yet. Up to 7 kV the current is mainly the ohmic current flowing through the surface of the gaps. Above 8 kV the avalanche of the gas ionization contributes to the exponential rise of the currents.

\begin{figure} [h]
\begin{center}
\includegraphics[width=9.0cm]{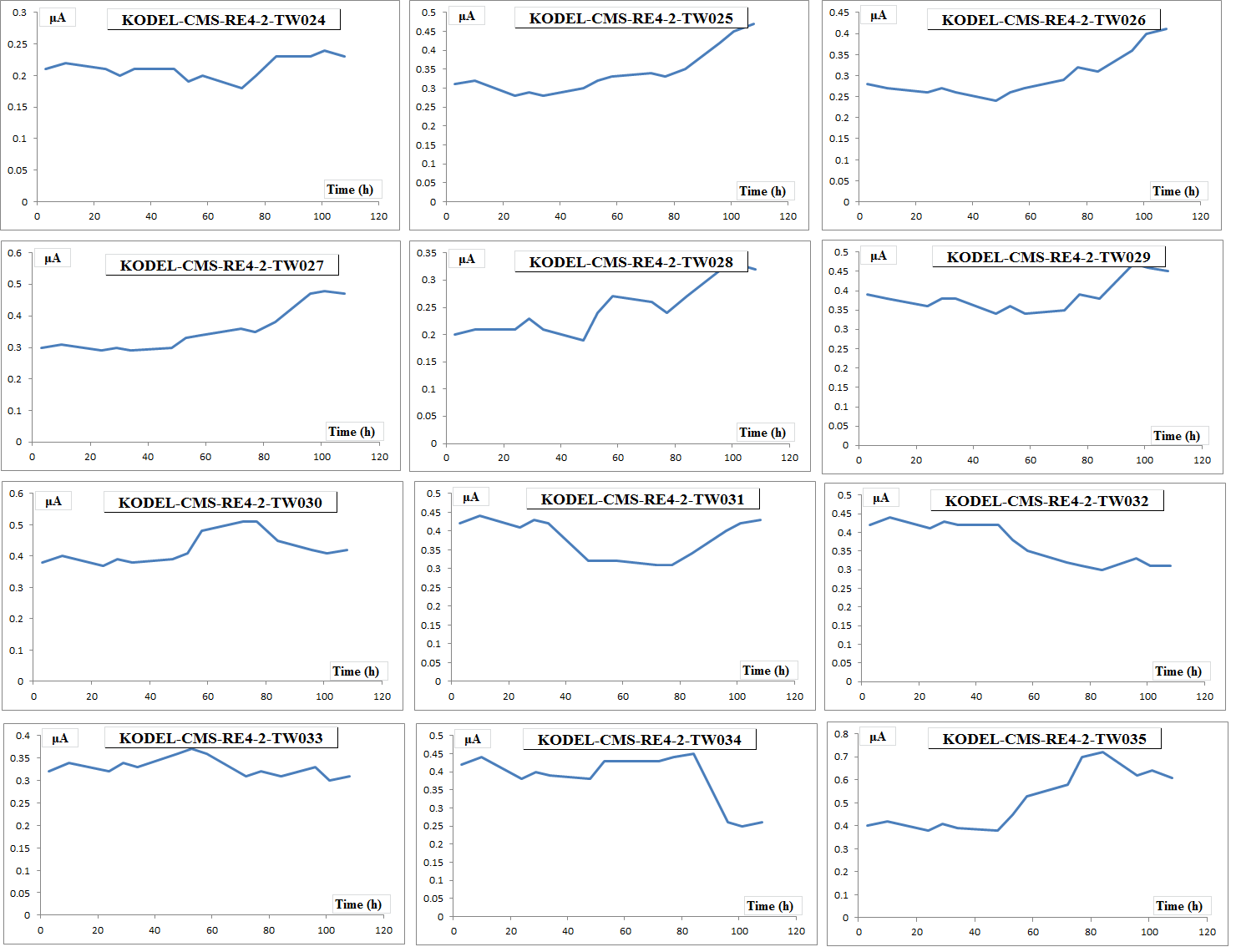}
\caption{Dark currents of gaps at 9.6 kV as a function of time up to 120 hours.}
\label{fig:dark_current96}
\end{center}
\end{figure}

A 120-hour long monitoring of gap currents at 9.6 kV is shown in Fig.~\ref{fig:dark_current96}. It shows the characteristic gradual decrease of the current during the initial 40 hours, and thereafter, it slightly increases.  
Finally, the H.V. scanning up to 10 kV is performed one more time. Two different measurements of the dark currents before and after the 120-hour long measurement are plotted against each other to detect any significant variations in Fig.~\ref{fig:dark_current10}.

\section{Resistivity of RPC Gap Measurement}

\begin{figure}[ht]
\begin{center}$
\begin{array}{cc}
\includegraphics[width=7cm]{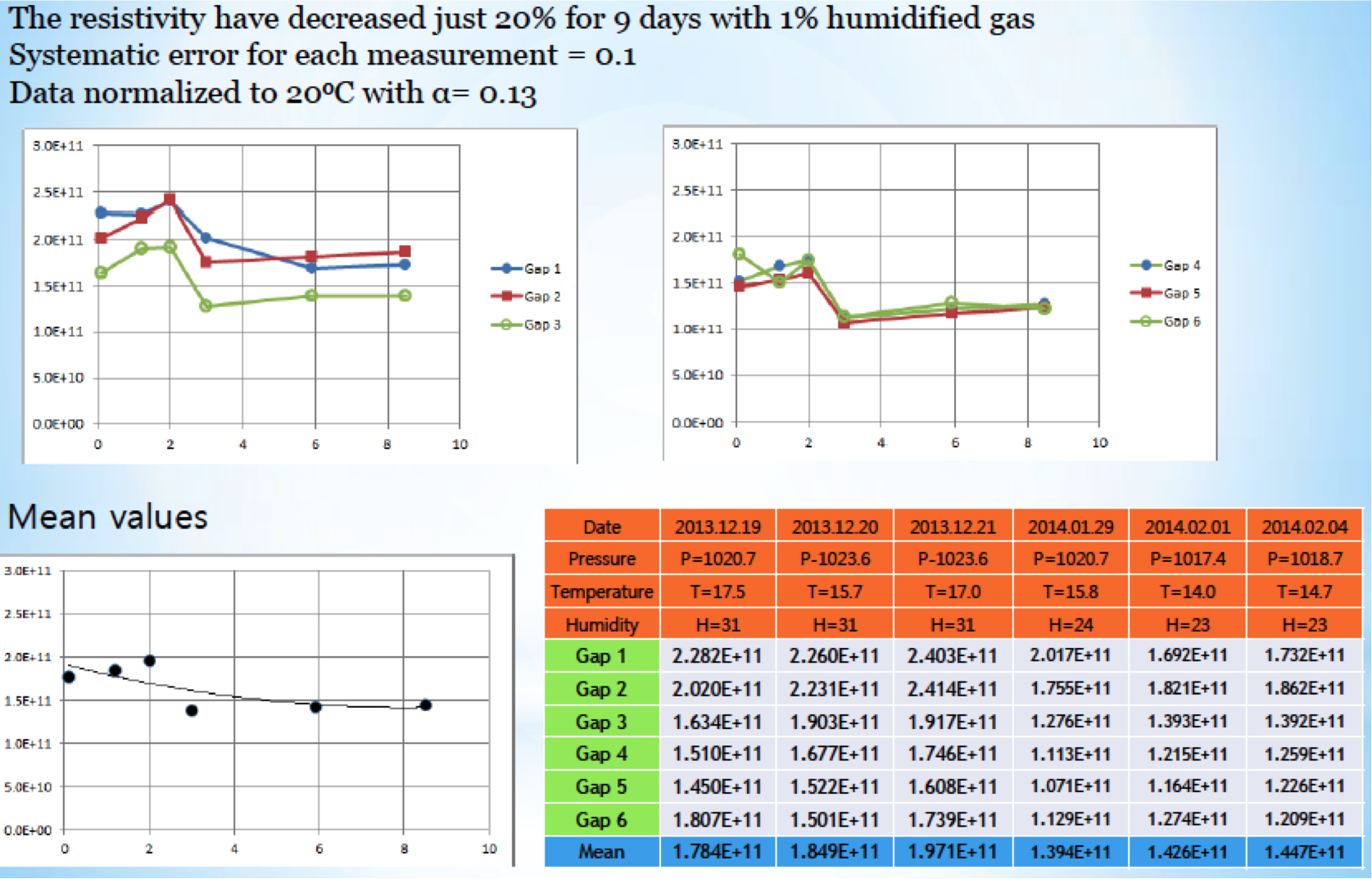} &
\includegraphics[width=4cm]{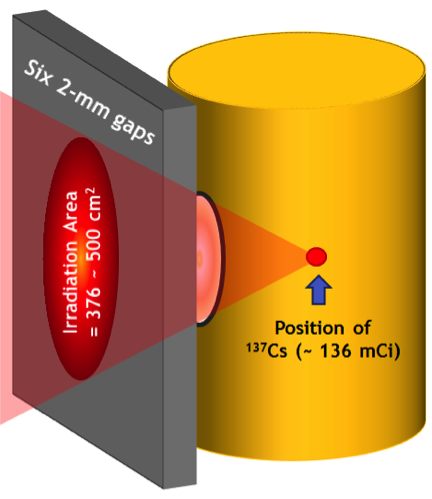}
\end{array}$
\end{center}
\caption{Resistivities of six small gaps (left) measured by gamma irradiations using $^{137}$Cs source (right). }
\label{fig:cs137}
\end{figure}

\begin{figure} [h]
\begin{center}
\includegraphics[width=9.0cm]{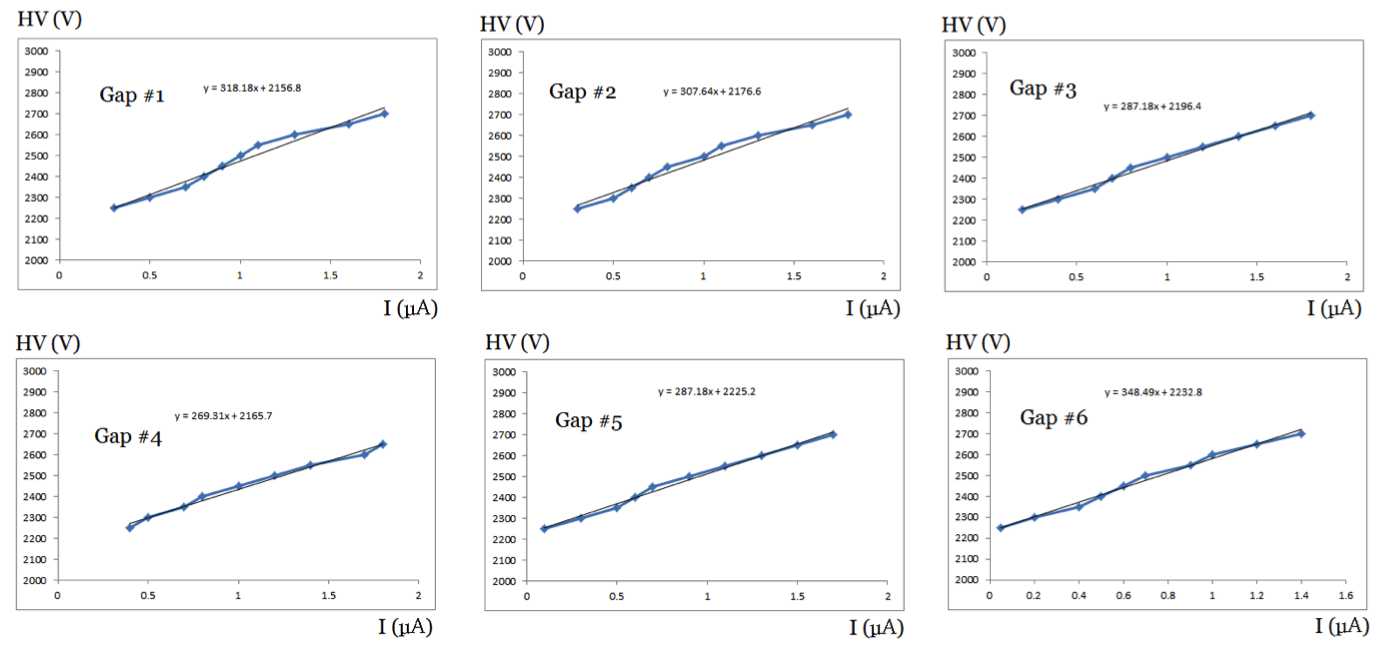}
\caption{Currents of the six small gaps as a function of H.V. with pure argon.}
\label{fig:argon}
\end{center}
\end{figure}

During QC-1 visual inspection, any damaged HPL panels were kept separately not to be used in the gap production. Out of some damaged HPL panels delivered to KODEL in May 2013, we produced six small RPC gaps. The effective area of each gap was 30~$\times$ 30~cm$^{2}$.  
We measured the resistivity of the RPC gaps by irradiating the six RPC gaps with 136-mCi~$^{137}$Cs source. The distance from gaps to the source
was between 26 and 35 cm as depicted in Fig.~\ref{fig:cs137}. The gas composition used in this test was Freon 95\% + Iosbutane 5\% with water vapor added with a mass ratio of 1\%.
As an independent measurement, we also used pure argon gas with water vapor added with a mass ratio of 1\%. Currents drawn from six small gaps during the measurement with pure argon as a function of H.V. are plotted in Fig.~\ref{fig:argon} 

The measured resistivities $\rho$, from both methods were normalized to the value at 20~$^{\circ}{\rm C}$.
$\rho_{b}^{20} = \rho_{T} e^{\alpha (T-T_0)} $, where the temperature coefficient $\alpha$=0.13 $^{\circ}{\rm C}^{-1}$.
From the $^{137}$Cs source irradiation we obtained the mean value of the resistivity of the six RPC gaps $\rho$=1.42$\pm$0.07 $\times$ 10${^{11}}\Omega$cm.
The resistivity measured using pure argon gas yielded the mean value of $\rho$=3.03 $\pm$ 0.23 $\times$ 10${^{11}}\Omega$cm.
The resistivity measured using pure argon gas was 2.13 times higher than the result obtained by the gamma irradiation method. 
Therefore, we drew an empirical conversion factor, ${\it C}$, to convert the resistivity from argon to the gamma irradiation ${\it C}$ = 0.47.

We then measured the resistivity of 23 TW RE4-2 gaps using the argon method. The yielded mean resistivity of the 23 TW gaps was $\rho$=1.35 $\times$ 10${^{11}}\Omega$cm. Applying the conversion factor obtained from the comparison of the irradiation with pure argon methods, we estimated the resistivity of the 23 TW4-2 gaps $\rho$ to be 6.33 $\times$ 10${^{10}}\Omega$cm. This exercise was to check the expected variation of the resistivity of the HPL panels over periods when stored without special climate control. From this frame of estimation, we found that the resistivity of the gaps was within the expected values.

\section{Conclusions} 


KODEL produced 686 RPC gaps in 22 months. The rate at which the qualified gaps are produced is 96\%.
These statistics show that the QC adopted in RPC gap production has been effective for the mass production of RPC gaps.

The resistivities of the RPC gaps were measured with two different methods; pure argon gas method and gamma irradiation method. 
We found that these two methods were comparable and the average resistivity of the gaps was within the expected values.

\acknowledgments
We acknowledge the support of Korea University and the Korean Research Foundation.

\end{document}